\documentstyle[m31x47jw]{book}
{\catcode `\| = 0 \catcode `\\ = 11 |long|gdef|filterouttext#1
\eskip{|endgroup}}

{\catcode `\| = 0 \catcode `\\ = 11 |gdef|bakslask{\}}
\def\eskip{\message{Warning: unmatched \bakslask eskip encountered}}

\textwidth=11.5cm 
\textheight=17.5cm 
\oddsidemargin=15mm
\topmargin=-5mm

\begin{document}

\bibliographystyle{plain}

\title{Chapter for Book Michael Wester}
\edition{First Edition, Wiley and Sons, New York (1999)}
\author{
Willy Hereman\footnote{
Department of Mathematical and Computer Sciences, 
Colorado School of Mines, Golden, CO 80401. 
Email: whereman@mines.edu
} 
and 
\"Unal G\"okta\c{s}\footnote{
Wolfram Research Inc., 100 Trade Center Drive, Champaign, IL 61820. 
Email: unalg@wolfram.com}
}


\booktitle{Computer Algebra Systems: A Practical Guide}


\maketitle
\tableofcontents

\chapter{Integrability Tests for Nonlinear Evolution Equations}
\chaptitle{Integrability Tests}

\section{Introduction}

During the last three decades, the study of integrability of nonlinear 
ordinary and partial differential equations (ODEs and PDEs) has been 
the topic of major research projects (see, e.g., \cite{MAandPC91,AMetal91}). 
This chapter presents a few symbolic algorithms to illustrate how 
computer algebra systems (CASs) can be effectively used in integrability
investigations. We work with {\it Mathematica} \cite{Wolfram96}, 
but our 
algorithms can be implemented in other languages. 

Among the many alternatives \cite{BGandAR97} for investigating the 
integrability of systems of PDEs with symbolic software, 
the search for conserved densities, generalized symmetries, 
and recursion operators is particularly appealing \cite{AMetal91}. 
Indeed, it turns out that these quantities can be computed 
without the use of sophisticated mathematical tools. 
As a matter of fact, not much beyond differentiation and solving of 
linear systems is needed. As a result, our algorithms are easy to 
implement. 
In fairness, our algorithms are restricted to the computation of
polynomial quantities of polynomial equations. Yet, this covers
the majority of the cases treated in the literature.

The algorithms in this chapter are based on a common principle: 
scaling (or dilation) invariance. 
Indeed, we observed that many known integrable systems are 
invariant under dilation symmetry, which is a special Lie point symmetry.
The dilation symmetry can be computed by solving a linear system. 
Using dilation invariance, the plan is to first produce candidates for the 
polynomial densities, symmetries, and recursion operators in an efficient 
way. Once the candidate expressions are computed, their unknown constant
coefficients follow from solving a linear system.  

We focus our attention on explaining the strategy, at the cost of 
mathematical rigor and details, which can be found in 
\cite{UG98,UGandWH97a,UGandWH98b,WHetal98}. 
Rather than discussing the algorithms in general, we apply them to a 
few prototypical nonlinear evolution equations from nonlinear wave theory. 
Whenever appropriate, we address issues related to the implementation of 
the algorithms. For instance, we give explicit code for the Fr\'echet 
derivative, which is one of the key tools in our methods. 

Our package {\it InvariantsSymmetries.m} \cite{UGandWH98c} 
works for nonlinear evolution equations. 
Applied to a system with parameters, our package can 
determine the conditions on the parameters so that the system admits a 
sequence of conserved densities or generalized symmetries.
Although we do not address it here, our package can also compute 
densities and symmetries of differential-difference equations 
(semi-discrete lattices). 
See \cite{UGandWH98a,UGandWH98b,UGetal97,WHetal98} for more information
about that subject. 

Due to memory constraints, our software can only compute a limited 
number of conserved densities and symmetries (half a dozen for
systems; at best a dozen for scalar equations). 
To prove integrability, one must show that infinitely many 
independent densities or symmetries exist. Alternatively, 
one could construct the operator that connects the 
symmetries, and prove that it is a true recursion operator.  
Such proofs involve mathematical methods \cite{AF87,AMetal91,Olver93,JPW98}
that are beyond the scope of this article.   
Although it is not yet implemented, we also present an algorithm 
for the computation of recursion operators, based on the knowledge 
of a few conserved densities and symmetries. 

The computation of Lie point symmetries and generalized symmetries 
via prolongation techniques is purposely omitted. 
That topic and related software were covered extensively in \cite{WH96,WH97}.
Space limitations also prevent the inclusion of the well-known 
Painlev\'e test, which is a widely applied and successful integrability 
detector for nonlinear ODEs and PDEs. We refer to \cite{WHetal98} 
for survey papers, books, and software related to Painlev\'e analysis. 

This chapter is organized as follows. 
In Section 2, we discuss scaling symmetries of PDEs and show how 
to compute them. 
Section 3 deals with conservation laws. We give the definition and
the steps of our algorithm, and show how to implement and apply 
the algorithm. We do the same for generalized symmetries in Section 4.
An algorithm to determine the recursion operator is given in Section 5. 
The leading examples in Sections 2 through 5 are the Korteweg-de Vries (KdV)
and Sawada-Kotera (SK) equations, and a system of nonlinear Schr{\"o}dinger-%
type equations. 
For the latter, we derive the recursion operator in Section 6. 
In Sections 7 and 8, we discuss our {\it Mathematica} package 
{\it InvariantsSymmetries.m} and review similar software. 
We draw some conclusions in Section 9.

\section{Key Concept: Dilation Invariance}

Our algorithms are based on the following observation: if a system of 
nonlinear evolution equations is invariant under a dilation (scaling) 
symmetry, then its conservation laws, generalized symmetries, 
and the recursion operator have the same scaling properties as the system.
This is at least true for the polynomial case.

As leading example, we use the ubiquitous Korteweg-de Vries (KdV) 
equation
\cite{MAandPC91}, 
\begin{equation}
\label{kdv}
u_t = 6 u u_x + u_{3x},
\end{equation} 
which describes water and plasma waves and lattice dynamics.
Throughout this chapter, we will use the notations 
\begin{equation} 
u_t = \frac{\partial u}{\partial t}, \quad
u_{tx} = \frac{\partial^2 u}{\partial t \; \partial x}, \quad
u_{nx} = \frac{\partial^n u}{\partial x^n}.
\end{equation}
Equation (\ref{kdv}) is invariant under the dilation (scaling) symmetry
\begin{equation}
\label{kdvdilation}
(t, x, u) \rightarrow (t / {\lambda}^{3}, x / \lambda, {\lambda}^{2} u),
\end{equation}
where $\lambda$ is an arbitrary parameter. 
Indeed, replacement of $(t,x,u)$ according to (\ref{kdvdilation})
allows one to cancel a factor $\lambda^5$ in (\ref{kdv}). 
Note that, e.g., $\frac{\partial}{\partial t}$ is replaced by 
$\lambda^3 \frac{\partial}{\partial t}.$ 
Obviously, $u$ corresponds to two derivatives in $x$, i.e., 
$u \sim {\partial^2}/{\partial {x^2}}. $
Similarly, ${\partial}/{\partial t} \sim {\partial}^3/{\partial x}^3.$ 

We express all scalings in terms of $\frac{\partial}{\partial x}.$ 
Introducing {\it weights}, denoted by $w,$ we could say that $w(u) = 2$ 
and $w({\rm D}_t) = 3,$ if we set $w({\rm D}_x) = 1.$ 
We used ${\rm D}_t$ and ${\rm D}_x$ instead of 
$w(\frac{\partial}{\partial t})$ and $ w(\frac{\partial}{\partial x})$ 
to cover cases where densities and symmetries depend explicitly on
$t$ and $x$ (see \cite{UGandWH98b,WHetal98} for examples).

The {\it rank} $R$ of a monomial is equal to the sum of all of its weights. 
Observe that (\ref{kdv}) is {\it uniform in rank} since all the terms 
have rank $R=5,$ confirming that $\lambda^5$ was a common factor.
\vskip 12pt
\noindent
{\bf Computation of scaling symmetries.}$\;\;$To compute the 
scaling symmetry of an equation, we compute the weights of all its
terms, and {\it require} that the equation be uniform in rank. 
For (\ref{kdv}), with $w({\rm D}_x) = 1,$ this yields 
\begin{equation}
w(u) + w({\rm D}_t) = 2 w(u) + 1 = w(u) + 3.
\end{equation}
The solution of this linear system is $w(u) = 2,$ and $w({\rm D}_t) = 3.$  
\vskip 8pt

As a second example, we consider a fifth-order PDE from soliton theory, 
\begin{equation}
\label{sk}
u_t = 5 u^2 u_x + 5 u_x u_{2x} + 5 u u_{3x} + u_{5x},  
\end{equation}
due to Sawada and Kotera \cite{KSandTK74}. 
Scaling invariance requires that   
\begin{equation}
\label{balancesk}
w(u) + w({\rm D}_t) = 3 w(u) + 1 = 2 w(u) + 3 = w(u) + 5.  
\end{equation}
Hence $w(u) = 2$ and $w({\rm D}_t) = 5.$ 
\vskip 10pt
\noindent
{\bf Systems.}$\;\;$Single PDEs like (\ref{kdv}) are a special case of 
\begin{equation}
\label{pdesys}
{\bf u}_t = {\bf F}({\bf u}, {\bf u}_{x}, {\bf u}_{2x}, \ldots, {\bf u}_{mx}), 
\end{equation}
where ${\bf u}$ and ${\bf F}$ are vector dynamical variables with $n$ 
components. The number of components, the order $m$ of the system, and 
its degree of nonlinearity are arbitrary. 

To determine the scaling symmetry, we require that each equation 
in (\ref{pdesys}) be uniform in rank, and solve the resulting linear 
system for the weights of all the variables. 

As an example, consider a vector nonlinear Schr{\"o}dinger equation, 
\begin{equation}
\label{DMVvector}
{\bf B}_t + (|{\bf B}|^2 {\bf B})_x + ({\bf B}_0 \cdot
{\bf B}_x) {\bf B}_0  + {\bf e} \times {\bf B}_{xx} = 0, 
\end{equation}
which occurs in plasma physics \cite{BDetal93a,BDetal93b}. 
With ${\bf B}_0 = (a,b)$ and ${\bf B} = (u,v)$ in the $(y,z)$-plane, 
and ${\bf e}$ along the $x$-axis, (\ref{DMVvector}) can be written as
\begin{eqnarray}
\label{DMV}
&& u_t + \left [ u ( u^2 + v^2 ) + \beta u + \gamma v - v_x \right ]_x = 0, 
\nonumber \\
&& v_t + \left [ v ( u^2 + v^2 ) + \theta u + \delta v + u_x \right ]_x = 0 ,  
\end{eqnarray}
where $\beta = a^2, \gamma = \theta = a b,$ and $\delta = b^2$ are 
nonzero parameters. 
With reference to \cite{BDetal93a}, we call (\ref{DMVvector})
or (\ref{DMV}) the DMV equation. 
To start generally, we will consider the system (\ref{DMV}) for arbitrary
nonzero parameters $\beta, \gamma, \theta $ and $\delta.$

System (\ref{DMV}) is not uniform in rank, unless we allow that the 
parameters $\beta$ through $\delta$ have weights. Doing so, 
with $w({\rm D}_x) = 1,$ we obtain
\begin{eqnarray}
\label{balanceDMV}
w(u) + w({\rm D}_t) 
\!\!&\!=\!&\!\! 
3 w(u) + 1 =  w(u) + 2 w(v) + 1 =  
w(u) + w(\beta) + 1 \nonumber \\
\!\!&\!=\!&\!\! 
w(v) + w(\gamma) + 1 = w(v) + 2, \\
w(v) + w({\rm D}_t) \!&\!=\!&\! 2 w(u) + w(v) + 1 = 3 w(v) + 1 
= w(u) + w(\theta) + 1
\nonumber \\
\!\!&\!=\!&\!\! 
w(v) + w(\delta) + 1 = w(u) + 2. 
\end{eqnarray}
Hence, 
\begin{equation}
\label{weightsDMV}
w(u) = w(v) = \frac{1}{2}, \;\; 
w(\beta) = w(\gamma) = w(\theta) = w(\delta) = 1, \;\; 
w({\rm D}_t) = 2.
\end{equation}
\vskip 10pt
\noindent
{\bf Remark.}$\;\;$For scaling-invariant equations like (\ref{kdv}) and 
(\ref{sk}), it suffices to consider the dilation symmetry on the space of 
independent and dependent variables. 
For systems like (\ref{DMV}) that are inhomogeneous for scaling, 
we give weights to the parameters to circumvent the problem. 
For systems that lack scaling invariance and have no parameters,
introducing one (or more) auxiliary parameter(s) with appropriate scaling
provides a solution. 

The trick is to extend the action of the dilation symmetry to the space 
of independent and dependent variables, {\it including} the parameters.
Doing so, our algorithms apply to a larger class of polynomial PDEs.
The extra parameters are only used in the first step of the algorithms: 
that is, in producing the candidate densities and generalized 
symmetries. Beyond that first step, parameters are no longer treated as 
dependent variables! Details and examples are given in 
\cite{UGandWH97a,UGandWH98b,WHetal98}.

\section{Conservation Laws}

{\bf Definition.}$\;\;$A conservation law for (\ref{pdesys}), 
\begin{equation}
\label{conslaw}
{\rm D}_{t} \rho + {\rm D}_{x} J = 0,  
\end{equation}
connects the {\em conserved density} $\rho$ and the 
{\em associated flux} $J.$ 
As usual, ${\rm D}_{t}$ and ${\rm D}_{x}$ are total derivatives, and 
(\ref{conslaw}) holds for all solutions of (\ref{pdesys}). 
Hence, density-flux pairs only depend on ${\bf u}, {\bf u}_x, $ 
etc., not on ${\bf u}_t.$ With a few exceptions, densities and fluxes 
do not explicitly depend on $t$ and $x.$ 

For the scalar case, $u_t = F,$ the computations are carried out 
as follows: 
\begin{equation}
\label{Dtrho}
{\rm D}_t \rho = \frac{\partial \rho}{\partial t} +
\sum_{k=0}^{n} \frac{\partial \rho}{\partial u_{kx}} {\rm D}_x^k u_t, 
\end{equation}
where $n$ is the order of $\rho.$
Upon replacement of $u_t, u_{xt},$ etc. from $u_t = F,$ one gets
\begin{equation}
\label{Dtrhonew}
{\rm D}_t \rho = \frac{\partial \rho}{\partial t} + \rho'(u) [F], 
\end{equation}
where $\rho'(u) [F]$ is the Fr\'echet derivative of $\rho$ in the 
direction of $F$ (see Section 1.4).
Furthermore, 
\begin{equation}
\label{DtJ}
{\rm D}_x J = \frac{\partial J}{\partial x} +
\sum_{k=0}^{m} \frac{\partial J}{\partial u_{kx}} u_{(k+1)x}, 
\end{equation}
where $m$ is the order of $J.$ 
Integrating both terms in (\ref{conslaw}) with respect to $x$ yields 
\begin{equation}
\label{}
{\rm D}_t \int_{-\infty}^{+\infty} \rho \; dx = 
- J |_{-\infty}^{+\infty} = 0, 
\end{equation}
provided that $J$ vanishes at infinity. In that case, 
\begin{equation}
\label{conservedP}
P = \int_{-\infty}^{+\infty} \rho \; dx = {\rm constant \; in \; time}.
\end{equation}
So, $P$ is the true conserved quantity. 
For ODEs, the quantities $P$ are called constants of motion. 
\vskip 8pt
\noindent
{\bf Examples.}$\;\;$The first three (of infinitely many) independent 
conservation laws \cite{MAandPC91,RMetal68} for (\ref{kdv}) are 
\begin{eqnarray} 
\label{kdvconslaw1}
\!\!\!\!\!&&\!\!\!\!
{\rm D}_t (u) - {\rm D}_x (3 u^2 + u_{2x} )=0, \\
\label{kdvconslaw2}
\!\!\!\!\!&&\!\!\!\!
{\rm D}_t (u^2) - {\rm D}_x ( 4 u^3 - u_x^2 + 2 u u_{2x} )=0,\\
\label{kdvconslaw3}
\!\!\!\!\!&&\!\!\!\!
{\rm D}_t \left(u^3 \!- \frac{1}{2} u_x^2\right) \!-
{\rm D}_x \left({9\over 2} u^4 \!- 6 u u_x^2 \!+ 3 u^2 u_{2x} 
\!+ \frac{1}{2} u_{2x}^2 \!- u_x u_{3x}\right)\!=0.
\end{eqnarray}
The first two conservation laws correspond to conservation of 
momentum and energy.
Note that the above conservation laws are indeed invariant under 
(\ref{kdvdilation}). 
The terms in the conservation laws have ranks $5, 7,$ and $9.$ 
The densities 
\begin{equation}
\label{conskdv}
\rho^{(1)} = u, \;\; \rho^{(2)} = u^2, \;\;\; {\rm and} \;\;\;
\rho^{(3)} = u^3 - \frac{1}{2} u_x^2 
\end{equation} 
have ranks $2, 4$ and $6$, respectively. 
The associated fluxes have ranks $4, 6$ and $8.$ 

Equation (\ref{kdv}) also has a density-flux pair that depends explicitly 
on $t$ and $x:$
\begin{eqnarray}
\label{xtrhofluxkdv}
{\tilde \rho} \!&\!=\!&\! t u^2 + \frac{1}{3} x u, \\
{\tilde J} \!&\!=\!&\! t (4 u^3 + 2 u u_{2x} - u_x^2) + 
x \left(u^2 + \frac{1}{3} u_{2x}\right) - \frac{1}{3} u_x.  
\end{eqnarray}
${\tilde \rho}$ has rank 1, ${\tilde J}$ has rank 3, since $w(t)=-3$
and $w(x) = -1.$
To accommodate this case, we used the total derivative notation
${\rm D}_t$ and ${\rm D}_x,$ in (\ref{conslaw}). 

For (\ref{sk}), the first two (of infinitely many) conserved densities 
\cite{UGandWH97a} are
\begin{equation} 
\label{conssk}
\rho^{(1)} = u \quad {\rm and} \quad 
\rho^{(2)} = {1 \over 3} u^3 - u_x^2.
\end{equation}
We will use them in the construction of the recursion operator 
for (\ref{sk}) in Section 5. 

We now describe how to compute {\it polynomial} densities that 
are (explicitly) independent of $t$ and $x.$
We refer to \cite{UG98,UGandWH97a} for the general algorithm covering 
systems as well as $(x,t)$-dependent densities. 
\vskip 8pt
\vskip 5pt
\noindent
{\bf Algorithm for Polynomial Conserved Densities}
\vskip 8pt
\noindent
{\bf Step 1: Determine the form of the density}
\vskip 5pt
\noindent
Select the rank $R$ of $\rho ,$ say, $R=6.$ 
Make a list ${\cal L}$ of all monomials in the components of ${\bf u}$ 
and their $x$-derivatives that have rank $R.$ 
Remove from ${\cal L}$ all monomials where the power of the highest 
derivative is $1.$ This is done to remove terms in $\rho$
that are in ${\rm Image}\; ({\rm D}_x),$ and therefore belong to the 
flux $J.$ 
After all, densities are equivalent if they only differ by terms that 
are total derivatives with respect to $x.$

Make a linear combination with constant coefficients $c_i$ of the monomials 
that eventually remain in the list ${\cal L}.$

For (\ref{kdv}), ${\cal L} = \{ u^3, u_x^2, u u_{2x}, u_{4x} \}.$ 
Next, $u_{4x}$ and $u u_{2x}$ are removed. Obviously, 
$u_{4x} = {\rm D}_x u_{3x},$  and $u u_{2x} = 
\frac{1}{2} {\rm D}_x^2 u^2 - u_x^2.$ So, $u u_{2x}$ and
$u_x^2$ only differ by a total $x$-derivative. From ${\cal L} = 
\{ u^3, u_x^2 \},$ one constructs $\rho = c_1 u^3 + c_2 u_x^2,$ 
which has rank $R=6.$
\vskip 8pt
\noindent
{\bf Step 2: Determine the unknown coefficients}
\vskip 4pt
\noindent
Substitute $\rho$ into the conservation law (\ref{conslaw}), and
compute ${\rm D}_t \rho$ via (\ref{Dtrho}). 
Use the PDE system to eliminate all $t$-derivatives of ${\bf u}$, 
and require the resulting expression $E$ to be a total $x$-derivative. 

To avoid integration by parts, apply the Euler operator 
(also called the variational derivative) \cite{Olver93} 
\begin{eqnarray}
\label{euleroperator}
L_u \!&\!=\!&\! 
\sum_{k=0}^{m} 
{(-{\rm D}_x)}^k \frac{\partial}{\partial u_{kx}} \nonumber \\
\!&\!=\!&\!\frac{\partial}{\partial{u}} -
{\rm D}_x \left(\frac{\partial}{\partial{{u_x}}}\right)+
{\rm D}_{x}^2 \left( \frac{\partial }{\partial{{u_{2x}}}}\right)+\cdots+
(-1)^m {\rm D}_{x}^m \left(\frac{\partial }{\partial{u_{mx}}}\right)
\end{eqnarray}
to $E$ of order $m.$ 
If $L_u (E) = 0$ immediately, then $E$ is a total $x$-derivative.
If $L_u (E) \ne 0,$ then the remaining expression must vanish 
identically. 
This yields a linear system for the constants $c_i.$ Solve the system. 
Carrying out these operations for (\ref{kdv}), 
one gets $c_1 = 1, c_2 = -\frac{1}{2}.$
\vskip 10pt
\noindent
{\bf Remark.}$\;\;$With (\ref{Dtrhonew}), the system for the $c_i$
follows from $L_u (\rho'(u) [F]) = 0$
by equating to zero the coefficients of monomials in $u$ and their 
$x$-derivatives. 
\vskip 10pt
\noindent
{\bf Implementation in Mathematica}
\vskip 4pt
\noindent
In {\it Mathematica}, $\verb|D|$ is the total derivative operator and 
the variational derivative (Euler operator) can be found in the 
Standard Add-on Package {\it Calculus`VariationalMethods`}.

For instance, returning to $\rho^{(3)}$ in (\ref{conskdv}), with 
(\ref{kdv}) and (\ref{Dtrho}), one computes
\begin{eqnarray}
\label{Dtrho3}
E\!&\!=\!&\!{\rm D}_t \rho^{(3)} = {\rho^{(3)}}'(u) [u_t] = 
(3 u^2 - u_x {\rm D}) u_{t} \nonumber \\
\!&\!=
\!&\!18 u^3 u_x - 6 u_x^3 - 6 u u_x u_{2x} 
+ 3 u^2 u_{3x} - u_x u_{4x}.
\end{eqnarray}
Application of the variational derivative,
\verb|VariationalD[E,u[x,t],{x,t}]| gives zero. 
That means that $E$ is a total $x$-derivative of a polynomial 
$J^{(3)}.$ 
Integration of $E = -{\rm D}_x J^{(3)}$ gives 
\begin{equation}
J^{(3)} = 
-({9 \over 2} u^4 -6 u u_x^2 +3 u^2 u_{2x} +\frac{1}{2} u_{2x}^2 
- u_x u_{3x}).
\end{equation}
\vskip 5pt
\noindent
{\bf Example.}$\;\;$With our package {\it InvariantsSymmetries.m}, 
we searched for conserved densities of (\ref{DMV}). 
Obviously, (\ref{DMV}) is a conservation law; thus,
$\rho^{(1)} = u$ and $\rho^{(2)} = v,$ without conditions on 
the parameters.
Additional conserved densities only exist if $\gamma = \theta.$
The first few are: 
\begin{eqnarray}
\label{DMVdens3}
\rho^{(3)}\!\!\!&\!=\!&\!\!u^2 + v^2 , \\
\label{DMVdens4}
\rho^{(4)}\!\!\!&\!=\!&\!\!{1 \over 2} ( u^2 + v^2 )^2 
+ ( \beta - \delta ) u^2 + 2 \theta u v + 2 v u_x , \\
\label{DMVdens5}
\rho^{(5)}\!\!\!&\!=\!&\!\!{1 \over 4} ( u^2 + v^2 )^3 
+ {1 \over 2} ( u_x^2 + v_x^2 ) + \theta u v ( u^2 + v^2 ) \nonumber \\
&& + {1 \over 4} ( \beta - \delta ) ( u^4 - v^4 ) 
+ 3 u^2 v u_x + v^3 u_x ,
\end{eqnarray}
and 
\begin{eqnarray}
\label{DMVdens6}
\rho^{(6)}\!\!\!&\!=\!&\!\!{5 \over 32} ( u^2 + v^2 )^4 
+ {3 \over 4} ( u^2 + v^2 ) ( u_x^2 + v_x^2 )
+ {1 \over 2} ( u u_x + v v_x )^2  \nonumber \\
&& + {1 \over 8} ( \beta - \delta )^2 u^4 
+ {1 \over 4} ( \beta - \delta ) u^6 
+ {1 \over 2} ( \beta - \delta ) \theta u^3 v
+ {3 \over 4} ( \beta - \delta ) \theta^2 v^2 \nonumber \\
&& + {3 \over 8} ( \beta - \delta ) u^4 v^2 
- {1 \over 4} \theta^2 ( u^4 + v^4 ) 
+ {3 \over 4} \theta ( u^5 v + u v^5) \nonumber \\
&& - {1 \over 8} ( \beta - \delta ) v^6 
- {3 \over 2} \theta^3 u v + {3 \over 2} \theta u^3 v^3
- {3 \over 2} \theta^2 v u_x \nonumber \\
&& + {3 \over 2} ( \beta - \delta ) u^2 v u_x 
+ {15 \over 4} u^4 v u_x + {3 \over 2} \theta u v^2 u_x 
+ {5 \over 2} u^2 v^3 u_x \nonumber \\
&& + {3 \over 4} v^5 u_x + {1 \over 4} ( \beta - \delta ) u_x^2 
+ {1 \over 2} \theta u_x v_x + {1 \over 2} v_x u_{2x}. 
\end{eqnarray}

Via integration by parts, $\rho^{(4)}$ is equivalent to
\begin{equation}
\label{rho4alt}
{\tilde \rho}^{(4)} = {1 \over 2} ( u^2 + v^2 )^2 + ( \beta - \delta ) u^2 
+ 2 \theta u v + v u_x - u v_x.
\end{equation}  
Indeed, $\rho^{(4)} = {\tilde \rho}^{(4)} + D_x (u v).$

Judged from (\ref{DMVdens3})--(\ref{DMVdens6}), the complexity of 
the expressions dramatically increases as the rank increases. 
The fact that we were able to compute 6 independent densities for 
(\ref{DMV}) is an indicator that the system presumably is completely 
integrable, as was later proved in \cite{RWetal95}.

Only the conserved densities $\rho^{(1)}$ through $\rho^{(3)}$ will be 
used in the construction of the recursion operator for (\ref{DMV}) 
in Section 6. 

\section{Generalized Symmetries}

{\bf Definition.}$\;\;$A vector function 
${\bf G} (x, t, {\bf u}, {\bf u}_{x}, {\bf u}_{2x}, \ldots)$ is 
called a {\it symmetry} of (\ref{pdesys}) if and only if it leaves 
(\ref{pdesys}) invariant under the replacement 
${\bf u} \rightarrow {\bf u} + \epsilon {\bf G}$ within order $\epsilon.$ 
Hence, 
\begin{equation}
\label{invariance}
{\rm D}_t ({\bf u} + \epsilon {\bf G}) = 
{\bf F} ({\bf u} + \epsilon {\bf G})
\end{equation}
must hold up to order $\epsilon$ on any solution of (\ref{pdesys}). 
Consequently, ${\bf G}$ must satisfy the linearized equation 
\cite{AF87,AMetal91}
\begin{equation}
\label{pdesymmetry}
{\rm D}_t {\bf G} = {\bf F}'({\bf u})[{\bf G}],
\end{equation}
where ${\bf F}'$ is the Fr\'echet derivative of {\bf F}, i.e.,
\begin{equation}
\label{pdefrechet}
{\bf F}'({\bf u})[{\bf G}] = 
{\partial \over \partial{\epsilon}} 
{\bf F}({\bf u}+\epsilon {\bf G})_{|{\epsilon = 0}} .
\end{equation}
\vskip 3pt
\noindent
In (\ref{invariance}) and (\ref{pdefrechet}), we infer that ${\bf u}$ 
is replaced by ${\bf u} + \epsilon {\bf G},$ and 
${\bf u}_{nx}$ by ${\bf u}_{nx} + \epsilon {\rm D}^n_x {\bf G}.$
As usual, ${\rm D}_{t}$ and ${\rm D}_{x}$ are total derivatives, 
and ${\bf G} =(G_1, G_2, \ldots, G_n)$ if the system 
(\ref{pdesys}) has $n$ components.

Once higher-order symmetries have been found, these vector fields can
be used to obtain fundamental information about the integrability of the 
equation. In many cases, conserved quantities, Hamiltonian structures, and
recursion operators follow readily from the knowledge of generalized 
symmetries \cite{AF87}.
\vskip 8pt
\noindent
{\bf Examples.}$\;\;$The first three (of infinitely many) symmetries 
\cite{Olver93} of (\ref{kdv}) are 
\begin{eqnarray}
\label{symkdv}
G^{(1)} \!&\!=\!&\! u_x, \;\;\;\;\; G^{(2)} = 6 u u_x + u_{3x}, \nonumber \\
G^{(3)} \!&\!=\!&\! 30 u^2 u_x + 20 u_x u_{2x} + 10 u u_{3x} + u_{5x}.
\end{eqnarray} 
All the terms in these symmetries have rank $3, 5$ and $7,$ respectively. 

With higher-order symmetries one can generate new integrable PDEs. 
For example, $u_t = G^{(3)}$ is the Lax equation in the completely 
integrable KdV hierarchy \cite{MAandPC91}. 

Note that (\ref{kdv}) also admits symmetries that explicitly 
depend on $t$ and $x.$ Indeed, the symmetries 
\begin{equation}
\label{xtsymkdv}
{\tilde G}^{(1)} = 1 + 6 t u_x \;\;\;{\rm and}\;\;\;
{\tilde G}^{(2)} = 4 u + 2 x u_x + 36 t u u_x + 6 t u_{3x}
\end{equation}
are of rank $0$ and $2.$ They linearly (and explicitly) depend on 
$t$ and $x.$

The algorithm presented in this paper can easily be extended to 
cover this type of symmetries (see \cite{UG98,UGandWH98b} for details).

The situation for (\ref{sk}), which also has infinitely many polynomial 
symmetries, is more complicated. 
The symmetries of (\ref{sk}) originate from two distinct ``seeds":
\begin{equation}
\label{symsk} 
G^{(1)} = u_x \quad {\rm and} \quad
G^{(2)} = 5 u^2 u_x + 5 u_x u_{2x} + 5 u u_{3x} + u_{5x}. 
\end{equation}
We have also computed symmetries of higher rank, but we do not show them here. 
A detailed computer-aided study showed that the symmetries $G^{(2i-1)}$ 
with rank $6i-3$ come from the seed $G^{(1)},$ whereas $G^{(2i)}$ with 
rank $6i+1$ originate from $G^{(2)},$ where $i=1,2,\ldots$
(see \cite{JSandJW98b} for details).  

For systems of type (\ref{pdesys}), the symmetry ${\bf G}$ is a vector 
with $n$ components. 
Our computer search with {\it InvariantsSymmetries.m} revealed 
that (\ref{DMV}) is invariant under the transformation 
$(u,v) \rightarrow (v,-u)$, which is a Lie point symmetry, 
provided the conditions $\beta = \delta $ and $ \gamma = - \theta$ hold. 
However, these conditions do not lead to a hierarchy of integrable equations. 
We therefore continued our search with arbitrary nonzero parameters
$\beta$ through $\delta.$

The first two symmetries of (\ref{DMV}) are
${\bf G}^{(1)} = (G_1^{(1)}, G_2^{(1)})$ and 
${\bf G}^{(2)} = (G_1^{(2)}, G_2^{(2)})$, where
\begin{eqnarray}
\label{DMVsymm1}
G_1^{(1)} \!&\!=\!&\! u_x , \nonumber \\
G_2^{(1)} \!&\!=\!&\! v_x , \\
\label{DMVsymm2}
G_1^{(2)} \!&\!=\!&\! ( \beta - \delta ) u_x + 3 u^2 u_x + v^2 u_x + 
\gamma v_x + 2 u v v_x - v_{2x} , \nonumber \\ 
G_2^{(2)} \!&\!=\!&\! \theta u_x + 2 u v u_x + u^2 v_x + 3 v^2 v_x + u_{2x}. 
\end{eqnarray}
Note that the sum of symmetries is still a symmetry. Remembering 
$(u_t, v_t) = (F_1, F_2)$ in (\ref{DMV}), we then have
$G_1^{(2)} + \delta G_1^{(1)} = -F_1$ and $ G_2^{(2)} = -F_2. $

The next symmetry, ${\bf G}^{(3)} = (G_1^{(3)}, G_2^{(3)}),$ 
only exists if $\gamma = \theta.$ It is
\begin{eqnarray}
\label{DMVsymm3}
G_1^{(3)} \!\!&\!\!\!=\!\!\!&\!\! 
3 ( \beta - \delta ) u^2 u_x + {15 \over 2} u^4 u_x 
+ 6 \theta u v u_x + 9 u^2 v^2 u_x 
+ {3 \over 2} v^4 u_x + 3 \theta ( u^2 + v^2 ) v_x 
\nonumber \\
&&\!+\!6 ( u^3 v + u v^3 ) v_x 
- 3 ( u^2 \!+ v^2 ) v_{2x} - 6 u u_x v_x - 6 v v_x^2 \!- u_{3x} , \\
G_2^{(3)} \!\!&\!\!\!=\!\!\!&\!\! 
3 \theta ( u^2 + v^2 ) u_x + 6 ( u^2 + v^2 ) u v u_x + 6 u u_x^2 
+ {3 \over 2} u^4 v_x + 6 \theta u v v_x \nonumber \\
&&\!+\!9 u^2 v^2 v_x - 3 ( \beta - \delta ) v^2 v_x + {15 \over 2} v^4 v_x 
+ 6 v u_x v_x + 3 ( u^2 \!+ v^2 ) u_{2x} - v_{3x}.
\end{eqnarray}
The algorithm for symmetries is similar to the one for conserved densities. 
The only difference is that monomials that differ by a total $x$-derivative 
are no longer removed from the list ${\cal L}$.
\vskip 15pt
\noindent
{\bf Algorithm for Polynomial Generalized Symmetries}
\vskip 10pt
\noindent
{\bf Step 1: Determine the form of the symmetry}
\vskip 6pt
\noindent
Select the rank $R$ of the symmetry. 
Make a list ${\cal L}$ of all monomials involving 
${\bf u}$ and its $x$-derivatives of rank $R.$ 
To obtain the form of the symmetry, make a linear combination of these 
monomials with constant coefficients $c_i.$
For example, for (\ref{kdv}), 
$G = c_1 \, u^2 u_x + c_2 \, u_x u_{2x} + c_3 \, u u_{3x} + c_4 \, u_{5x}$
is the form of the generalized symmetry of rank $R=7.$ 
\vskip 10pt
\noindent
{\bf Step 2: Determine the unknown coefficients}
\vskip 6pt
\noindent
Compute ${\rm D}_t {\bf G}.$ 
Use the PDE system to remove all $t$-derivatives. 
Equate the result to the Fr\'echet derivative ${\bf F}'({\bf u})[{\bf G}].$
Treat the different monomial terms in ${\bf u}$ and its $x$-derivatives 
as independent to get the linear system for the $c_i.$ Solve that system. 
For (\ref{kdv}), one obtains the symmetry of rank $7:$
\begin{equation}
\label{laxsymmagain}
G =  30 u^2 u_x + 20 u_x u_{2x} + 10 u u_{3x} + u_{5x}. 
\end{equation}
Symmetries of lower or higher rank are computed similarly.
See \cite{UG98,UGandWH98b,WHetal98} for details about the algorithm 
and its implementation. 
\vskip 12pt
\noindent
{\bf Remark.}$\;\;$ Starting with a conserved density $\rho$, 
the symmetries for a Hamiltonian system can be obtained from 
${\rm D}_x (L_u (\rho )),$ where $L_u$ is defined in (\ref{euleroperator}). 
See, for example, \cite{VRandGK94} for a study of the connection 
between densities and symmetries for Lagrangian and non-Lagrangian systems.
\vskip 15pt
\noindent
{\bf Implementation in Mathematica}
\vskip 5pt
\noindent
The key tool to compute symmetries is the Fr\'echet derivative, 
which is implemented as follows:
\vskip 4pt
{\hfuzz=1pt
\noindent
\begin{verbatim}
frechet[funcF_List,funcU_List,indVars_List,funcG_List]:=
Module[{eps,resultlist={},i},
Do[resultlist=Append[resultlist,
 Expand[(D[Part[funcF,i] /. 
 {(f_/; MemberQ[funcU,f])[Sequence @@ indVars]:>
 f[Sequence @@ indVars]+eps*funcG[[Flatten[Position[funcU,f]][[1]]]],
 Derivative[k_,0][f_][Sequence @@ indVars]:>
 D[f[Sequence @@ indVars]+eps*funcG[[Flatten[Position[funcU,f]][[1]]]],
 {indVars[[1]],k}]},eps]/. eps :> 0)]],
 {i,Length[funcF]}]; 
Return[resultlist]];
\end{verbatim}
}
\vskip 3pt
\noindent
To compute the Fr\'echet derivative of $u^3 - \frac{1}{2} u_x^2$ in 
the direction of $k(x,t),$ type
\begin{verbatim}
frechet[{u[x,t]^3-(1/2)*D[u[x,t],x]^2},{u},{x,t},{k[x,t]}];
\end{verbatim}
This gives the answer $3 u^2 k - u_x k_x.$
\section{Recursion Operators for Scalar Equations}

{\bf Definition.}$\;\;$A {\it recursion operator} is a linear
operator ${\bf \Phi}$ on the space of differential functions with 
the property that whenever $\bf G$ is a symmetry of (\ref{pdesys}), so 
is $\bf \hat{G}$ with $ {\bf {\hat{G}}} = {\bf \Phi} {\bf G}.$

The equation for the recursion operator \cite{AF87,Olver93,JPW98} is
\begin{equation}
\label{recursion}
{\rm D}_t {\bf \Phi} + [{\bf \Phi}, {\bf F}'(u) ] = 
\frac{\partial {\bf \Phi}}{\partial t} + 
{\bf \Phi}' [{\bf F}] + {\bf \Phi} \circ {\bf F}'(u) - 
{\bf F}'(u) \circ {\bf \Phi} = 0, 
\end{equation}
where $[\; , \; ]$ means commutator, $\circ$ indicates for composition, 
and the variational derivative of the operator ${\bf \Phi}$ is 
defined in, e.g., \cite{AF87}. 

For $n$-component systems like (\ref{pdesys}), the symmetries ${\bf G}$ 
are vectors with $n$ components, and the recursion operator ${\bf \Phi}$ 
is an $n \times n$ matrix.
\vskip 5pt
\noindent
{\bf Examples.}$\;\;$The recursion operator for (\ref{kdv}) is
\begin{equation}
\label{opkdv}
\Phi_{\rm KdV} = {\rm D}^2 + 4 u + 2 u_x {\rm D}^{-1}
     = {\rm D}^2 + 2 u + 2 {\rm D} u {\rm D}^{-1} \;,
\end{equation}
where, for simplicity of notation, 
${\rm D}={\rm D}_x$ and ${\rm D}^{-1}={\rm D}_x^{-1}.$ 
Here, $\Phi_{\rm KdV}\, G^{(1)} = G^{(2)}$ and 
$\Phi_{\rm KdV}\, G^{(2)} = G^{(3)}$
for the symmetries listed in (\ref{symkdv}). 
The recursion operator $\Phi_{\rm KdV}$ also connects the
$(x,t)$-dependent symmetries (\ref{xtsymkdv}), i.e., 
$\Phi_{\rm KdV} {\tilde G}^{(1)} = {\tilde G}^{(2)},$
but $\Phi_{\rm KdV} {\tilde G}^{(2)}$ is no longer local. 

Since $w({\rm D}^{-1})=-w({\rm D})= -1,$ the three terms in (\ref{opkdv}) 
have rank $R=2.$ The recursion operator (\ref{opkdv}) is uniform in rank.
Clearly, the rank of $\Phi_{\rm KdV}$ is the difference in rank between 
consecutive symmetries in (\ref{symkdv}).

In view of the symmetries (\ref{symsk}), the recursion operator 
for (\ref{sk}) must have rank $6.$ 
Indeed, the recursion operator \cite{BFandWO1982,BFetal87} has rank $6:$
\begin{eqnarray}
\label{opsk}
\Phi_{\rm SK}\!\!\!&\!=\!&\!\!\!\!
{\rm D}^6 + 2 u {\rm D}^4 + 2 {\rm D} u {\rm D}^3
+ {\rm D}^2 u {\rm D}^2 + 3 u {\rm D} u {\rm D} + 3 u {\rm D}^2 u 
- 2 {\rm D} u {\rm D} u - 2 u^3 \nonumber \\
\!\!\!\!&&\!\!\!\!+
{\rm D}^5 u {\rm D}^{-1} + 5 {\rm D} u {\rm D}^2 u {\rm D}^{-1} 
+ 5 u^2 {\rm D} u {\rm D}^{-1} + {\rm D} u {\rm D}^{-1} (u^2 - 2 u_x {\rm D}),
\end{eqnarray}
which can also be written as 
\begin{eqnarray}
\label{finalopsk}
\Phi_{\rm SK}\!\!\!&\!=\!&\!\!\!\!
{\rm D}^6 + 3 u {\rm D}^4 - 3 {\rm D} u {\rm D}^3
+ 11 {\rm D}^2 u {\rm D}^2 - 10 {\rm D}^3 u {\rm D} 
+ 5 {\rm D}^4 u \nonumber \\
\!\!\!\!&&\!\!\!\!+ 
12 u^2 {\rm D}^2 - 19 u {\rm D} u {\rm D}  + 8 u {\rm D}^2 u 
+ 8 {\rm D} u {\rm D} u + 4 u^3 \nonumber \\
\!\!\!\!&&\!\!\!\!+
u_x {\rm D}^{-1} (u^2 - 2 u_x {\rm D}) + G^{(2)} {\rm D}^{-1}, 
\end{eqnarray}
with $G^{(2)}$ in (\ref{symsk}). 

Our algorithm for the computation of polynomial recursion operators 
is based on the following observations.
\vskip 10pt
\noindent
{\bf Key observations.}$\;\;$All terms in (\ref{opkdv}) and 
(\ref{finalopsk}) are monomials in ${\rm D},{\rm D}^{-1}, u,$ and $u_x.$
Depending on the form of the recursion operator, $u_{2x}, u_{3x}, $ etc. 
can also appear, as is the case in (\ref{opsk}). 

Recursion operators split naturally in ${\Phi} = {\Phi}_0 + {\Phi}_1, $
where ${\Phi}_0$ is a differential operator (without ${\rm D}^{-1}$ 
terms), and ${\Phi}_1$ is an integral operator (with ${\rm D}^{-1}$ terms).

Furthermore, application of $\Phi$ to any symmetry should not leave 
any integrals unresolved, since all symmetries are polynomial
(see \cite{JSandJW97c}).
This is where the connection between conserved densities and symmetries 
comes into play. 

For instance, for (\ref{kdv}) it is clear that 
${\rm D}^{-1} (6 u u_x + u_{3x}) = 3 u^2 + u_{2x}$ is polynomial.
Similarly, for (\ref{sk}), using (\ref{finalopsk}), we have  
\begin{equation}
\label{skobs1}
{\rm D}^{-1} (5 u^2 u_x + 5 u_x u_{2x} + 5 u u_{3x} + u_{5x} ) 
= {5 \over 3} u^3 + 5 u u_{2x} + u_{4x}
\end{equation}
and
\begin{eqnarray}
\label{skobs2}
&& {\rm D}^{-1} ( u^2 - 2 u_x  {\rm D}) 
(5 u^2 u_x + 5 u_x u_{2x} + 5 u u_{3x} + u_{5x} ) \nonumber \\
&& \; = u^5 - 10 u^2 u_x^2 + \cdots - 2 u_x u_{5x}.
\end{eqnarray}
The first two conserved densities of (\ref{sk}) are $\rho^{(1)} = u$ 
and $\rho^{(2)} = {1 \over 3} u^3 - u_x^2$. 
Thus, with (\ref{conslaw}), we get 
$ {\rm D}_t u = u_t = - {\rm D}_x J^{(1)} $ and
\begin{equation}
\label{piecessk}
{\rm D}_t \left({1 \over 3} u^3 - u_x^2\right) = \rho'(u) [u_t] 
= (u^2 - 2 u_x {\rm D}) u_t = - {\rm D}_x J^{(2)}.
\end{equation}
So, the factor $(u^2 - 2 u_x {\rm D})$ in (\ref{finalopsk}) comes from 
$\rho^{(2)},$ and ${\rm D}^{-1} [(u^2 - 2 u_x {\rm D}) u_t]$ will 
be polynomial, namely $-J^{(2)}.$  

A similar situation happens for (\ref{kdv}), where 
$\rho^{(1)} = u, \rho^{(2)} = u^2,$ and 
$\rho^{(3)} = u^3 - {1 \over 2} u_x^2.$ 
Then, with (\ref{conslaw}) and (\ref{Dtrhonew}),
\begin{eqnarray}
\label{pieceskdv}
\!\!\!\!\!\!\!\!\!
&&\!\!\!
{\rm D}_t \rho^{(1)} = {\rm D}_t u = u_t = - {\rm D}_x J^{(1)}, \quad 
{\rm D}_t \rho^{(2)} = {\rm D}_t u^2 = 
2 u u_t = - {\rm D}_x J^{(2)}, \;\;\; {\rm and} \nonumber \\
\!\!\!\!\!\!\!\!\!
&&\!\!\!
{\rm D}_t \rho^{(3)} = {\rm D}_t ( u^3 - {1 \over 2} u_x^2 ) 
= {\rho^{(3)}}' (u) [u_t] = 
(3 u^2 - u_x {\rm D}_x ) u_t = - {\rm D}_x J^{(3)} ,
\end{eqnarray}
for polynomial $J^{(i)}, \; i=1,2,3.$
Thus, application of ${\rm D}^{-1},$ or ${\rm D}^{-1} u,$ or
${\rm D}^{-1} (3 u^2 - u_x {\rm D})$ to $ 6 u u_x + u_{3x}$ 
leads to a polynomial result.
However, as will be shown below, the terms involving ${\rm D}^{-1} u$ 
and ${\rm D}^{-1} (3 u^2 - u_x {\rm D})$ are not needed in the 
construction of the recursion operator (\ref{opkdv}). 

Since our algorithm for recursion operators is the most elaborate, we give
the steps that lead to (\ref{opkdv}) and (\ref{finalopsk}). We consider the 
scalar case first. Systems are dealt with in Section 6.  
\vskip 15pt
\noindent
{\bf Algorithm for Polynomial Recursion Operators}
\vskip 8pt
\noindent
{\bf Step 1: Construct the form of the recursion operator}
\vskip 7pt
\noindent
{\bf (i) Determine the rank of the operator}
\vskip 3pt
\noindent
Compute the rank $R$ of the operator based on the known ranks of 
consecutive symmetries.
For example, from (\ref{symkdv}), we compute
\begin{equation}
\label{ranksymkdv}
R = {\rm rank} \; \Phi = 
{\rm rank}\; G^{(3)} - {\rm rank}\; G^{(2)} 
= {\rm rank}\; G^{(2)} - {\rm rank}\; G^{(1)} = 2. 
\end{equation}
Obviously, 
${\rm rank} \; {\Phi}_0 = {\rm rank} \; {\Phi}_1 = {\rm rank} \; {\Phi} = R.$
\vskip 8pt
\noindent
{\bf (ii) Determine the pieces of the operator ${\Phi}_0$}
\vskip 3pt
\noindent
Make a list ${\cal L}$ of all permutations of ${\rm D}^j u^k,$ 
with $j$ and $k$ nonnegative integers, 
that have the rank $R.$ For (\ref{kdv}), 
\begin{equation}
\label{permkdv}
{\cal L} = \{ {\rm D}^2, u \}.
\end{equation}
\vskip 12pt
\noindent
{\bf (iii) Determine the pieces of the operator ${\Phi}_1$}
\vskip 2pt
\noindent
It can be shown \cite{AB93,JPW98} that 
\begin{equation}
\label{phi1part}
{\Phi}_1 = \sum_j \sum_k G^{(j)} {\rm D}^{-1} {\rho^{(k)}}'(u),    
\end{equation}
where the symmetries $G^{(j)}$ are combined with ${\rm D}^{-1}$ and
${\rho^{(k)}}'(u)$ in such a way that every term is exactly of 
${\rm rank}\; {\Phi}_1 = R.$
That is, the indices $j$ and $k$ are taken so that 
${\rm rank}\; (G^{(j)}) + {\rm rank}\;({\rho^{(k)}}'(u)) - 1 = R$ 
for every term in (\ref{phi1part}). 

Using the densities and symmetries, make a list ${\cal M}$ of 
the pieces involving ${\rm D}^{-1}.$ 
\vskip 0.001pt
\noindent
\begin{table}
\vspace{10pt}
\caption{{\rm Building blocks of ${\Phi}_1$ for the KdV equation.}}
\label{HGtable1}
\vskip .001pt
\noindent
\begin{center}
\begin{tabular}{| l | l | l | l |} \hline 
{\rule[-3mm]{0mm}{8mm} Rank} 
& Symmetry $G^{(j)}$& Density $\rho^{(k)}$ & ${\rho^{(k)}}'(u)$
 \\ \hline
& & & \\
0 & --- & --- & $1$ \\
1 & --- & --- & --- \\
2 & --- & $u$ & $u$ \\
3 & $u_x$ & --- & --- \\
4 & --- & $u^2$ & $3 u^2 - u_x {\rm D}$\\
5 & $6 u u_x + u_{3x}$ & --- & --- \\
6 & --- & $u^3 - \frac{1}{2} u_x^2$ & \\
& & & \\
\hline 
\end{tabular}
\end{center}
\end{table}
\vskip .001pt

For (\ref{kdv}), from Table~\ref{HGtable1} it should be clear that 
${\rm D}^{-1}$ can only be sandwiched between $u_x$ and $1.$ 
Any other combination would exceed rank $2$. 
Hence,
\begin{equation}
\label{mforkdv} 
{\cal M} = \{ u_x {\rm D}^{-1} \}. 
\end{equation}
\vskip 5pt
\noindent
{\bf (iv) Build the operator ${\Phi}$}
\vskip 2pt
\noindent
Next, produce ${\cal R} = {\cal L} \bigcup {\cal M},$ 
which has the building blocks of the recursion operator.
To get $\Phi,$ linearly combine the pieces in ${\cal R}$ with 
constant coefficients $c_i.$ For (\ref{kdv}), we obtain 
\begin{equation}
\label{rforkdv}
{\cal R} = \{ {\rm D}^2, u, u_x {\rm D}^{-1} \}. 
\end{equation}
Thus,
\begin{equation}
\label{formopkdv}
\Phi_{\rm KdV} = 
c_1 \, {\rm D}^2 + c_2 \, u + c_3 \, u_x {\rm D}^{-1}.
\end{equation}
\vskip 6pt
\noindent
We now repeat steps (i)--(iv) for the SK equation (\ref{sk}).
\vskip 2pt
\noindent
(i) Using the symmetries (\ref{symsk}), we get
\begin{equation}
\label{ranksymsk}
{\rm rank}\; \Phi = 
{\rm rank}\; G^{(3)} - {\rm rank}\; G^{(1)} = 
{\rm rank}\; G^{(4)} - {\rm rank}\; G^{(2)} = 6. 
\end{equation}
(ii) The operator ${\Phi}_0$ will be built from 
\begin{eqnarray}
\label{permsk}
{\cal L} \!&\!=\!&\! \{ {\rm D}^6 , u {\rm D}^4 , {\rm D} u {\rm D}^3 ,
{\rm D}^2 u {\rm D}^2 , {\rm D}^3 u {\rm D} ,
{\rm D}^4 u , u^2 {\rm D}^2 , u {\rm D} u {\rm D}, \nonumber \\ 
&& u {\rm D}^2 u, {\rm D} u^2 {\rm D}, {\rm D} u {\rm D} u, 
{\rm D}^2 u^2, u^3 \}.
\end{eqnarray}
\begin{table}[h] 
\caption{{\rm Building blocks of ${\Phi}_1$ for the SK equation.}}
\label{HGtable2}
\vskip 0.001pt
\noindent
\begin{center}
\begin{tabular}{| l | l | l | l |} \hline 
{\rule[-3mm]{0mm}{8mm}
Rank} & Symmetry $G^{(j)}$ & Density $\rho^{(k)}$ & 
${\rho^{(k)}}'(u)$ \\ \hline
& & & \\
0 & --- & --- & $1$ \\
1 & --- & --- & --- \\
2 & --- & $u$ & --- \\
3 & $u_x$ & --- & --- \\
4 & --- & --- & $u^2 - 2 u_x {\rm D}$\\
5 & --- & --- & --- \\
6 & --- & $\frac{1}{3} u^3 - u_x^2$ & --- \\
7 & $5 u^2 u_x + 5 u_x u_{2x} + 5 u u_{3x} + u_{5x} $ & --- & --- \\
& & & \\
\hline 
\end{tabular}
\end{center}
\end{table}
\vskip 0.001pt
\noindent
(iii) From Table~\ref{HGtable2}, which list the building blocks for ${\Phi}_1$ 
for (\ref{conssk}), we obtain
\begin{equation}
\label{integsk}
{\cal M} = \{ u_x {\rm D}^{-1} (u^2 - 2 u_x {\rm D}), 
(5 u^2 u_x + 5 u_x u_{2x} + 5 u u_{3x} + u_{5x}) {\rm D}^{-1} \}.
\end{equation}

All other combinations of the form $ G^{(j)} {\rm D}^{-1} {\rho^{(k)}}'(u)$ 
exceed rank $6.$
\vskip 5pt
\noindent
(iv) Combining the monomials from ${\cal R} = {\cal L} \bigcup {\cal M},$
we get
\begin{eqnarray}
\label{formopsk}
\Phi_{\rm SK}\!&\!=\!&\! 
c_1 {\rm D}^6 + c_2 u {\rm D}^4 + c_3 {\rm D} u {\rm D}^3
+ c_4 {\rm D}^2 u {\rm D}^2 + c_5 {\rm D}^3 u {\rm D}
\nonumber \\ 
\!&\!\!&\!+ 
c_6 {\rm D}^4 u + c_7 u^2 {\rm D}^2 + c_8 u {\rm D} u {\rm D} 
+ c_9 u {\rm D}^2 u + c_{10} {\rm D} u^2 {\rm D} \nonumber \\ 
\!&\!\!&\!+
c_{11} {\rm D} u {\rm D} u + 
c_{12} {\rm D}^2 u^2 + c_{13} u^3 + 
c_{14} u_x {\rm D}^{-1} (u^2 - 2 u_x {\rm D}) \nonumber \\
\!&\!\!&\!+
c_{15} (5 u^2 u_x + 5 u_x u_{2x} + 5 u u_{3x} + u_{5x}) {\rm D}^{-1} .
\end{eqnarray}
\vskip 8pt
\noindent
{\bf Step 2: Determine the unknown coefficients}
\vskip 4pt
\noindent
To determine the coefficients $c_i$, require that 
\begin{equation}
\label{connection}
\Phi G^{(k)} = G^{(k+s)}, \quad k = 1, 2, 3, \ldots ,
\end{equation}
where $s$ is the number of seeds. In practice, it suffices to use 
$k=1$ and $2$ in (\ref{connection}) to fix all coefficients $c_i.$
Solve the resulting linear system(s) for the unknown $c_i.$

For (\ref{kdv}), $\Phi_{\rm KdV} G^{(2)} = G^{(3)}$ with $\Phi_{\rm KdV}$ 
in (\ref{formopkdv}) and its symmetries (\ref{symkdv}), we obtain
\[
{\cal S}=\{ c_1 - 1 = 0, 18 c_1 + c_3 - 20 = 0, 
6 c_1 + c_2 - 10 = 0, 2 c_2 + c_3 - 10 = 0 \}.   
\]
The solution is 
$ c_1  = 1, c_2 = 4,$ and $c_3 = 2 $. 
Substituting it into (\ref{formopkdv}), we get
\begin{equation}
\label{finalopkdv}
\Phi_{\rm KdV} = 
{\rm D}^2 + 4 \, u + 2 \, u_x {\rm D}^{-1}. 
\end{equation}
The explicit computation on page 260 in \cite{AF87}
shows that (\ref{finalopkdv}) satisfies (\ref{recursion}). 

For (\ref{sk}), we express that
\begin{equation}
\label{connsk} 
\Phi_{\rm SK} G^{(1)} = G^{(3)} \quad {\rm and} \quad 
\Phi_{\rm SK} G^{(2)} = G^{(4)},
\end{equation}
with $\Phi_{\rm SK}$ in (\ref{formopsk}) and the symmetries (\ref{symsk}). 
Then we solve for the constants $c_i.$ This eventually yields 
\begin{eqnarray}
\label{ouropsk}
\Phi_{\rm SK}\!\!\!&\!=\!&\!\!\!\!
{\rm D}^6 + 3 u {\rm D}^4 - 3 {\rm D} u {\rm D}^3
+ 11 {\rm D}^2 u {\rm D}^2 - 10 {\rm D}^3 u {\rm D} 
+ 5 {\rm D}^4 u \nonumber \\
\!\!\!\!&&\!\!\!\!+
12 u^2 {\rm D}^2 - 19 u {\rm D} u {\rm D}  + 8 u {\rm D}^2 u 
+ 8 {\rm D} u {\rm D} u + 4 u^3 \nonumber \\
\!\!\!\!&&\!\!\!\!+
u_x {\rm D}^{-1} (u^2 - 2 u_x {\rm D}) + G^{(2)} {\rm D}^{-1}, 
\end{eqnarray}
with $G^{(2)}$ in (\ref{symsk}). 
A lengthy computation shows that this recursion operator satisfies 
(\ref{recursion}).

After integration by parts, (\ref{opsk}) or, equivalently, (\ref{ouropsk}) 
can also be written \cite{JSandJW98b} as
\begin{eqnarray}
\label{equivopsk}
\!\!\!\!\Phi_{\rm SK} \!\!\!&\!=\!&\!\!\! 
{\rm D}^6 + 6 u {\rm D}^4 + 9 u_x {\rm D}^3 
+ 9 u^2 {\rm D}^2 + 11 u_{2x} {\rm D}^2 + 10 u_{3x} {\rm D}
+ 21 u u_x {\rm D}  \nonumber \\
\!\!\!&\!\!&\!\!+
4 u^3 \!+ 16 u u_{2x} \!+ 6 u_x^2 \!+ 5 u_{4x} 
\!+ u_x {\rm D}^{-1} ( u^2 \!+ 2 u_{2x}) \!+ G^{(2)} {\rm D}^{-1}. 
\end{eqnarray}
\section{Recursion Operators for Systems}

We show how to construct the recursion operators for 
systems (\ref{pdesys}) with $n$ components. 
The symmetry ${\bf G}$ has $n$ components and the recursion 
operator ${\bf \Phi}$ is a $n \times n$ matrix. 

We used {\it Mathematica} interactively to compute the recursion operator 
for (\ref{DMV}) with $\gamma = \theta.$ 
The recursion operator has the form 
\begin{equation}
\label{opDMV}
{\bf \Phi} =
\left (
   \begin{array}{cc}
    {\Phi}_{11} & {\Phi}_{12} \\
    {\Phi}_{21} & {\Phi}_{22}
   \end{array}
\right),  
\end{equation}
>From ${\bf G}^{(2)} = {\bf \Phi} {\bf G}^{(1)}$, the rank of the 
entries ${\Phi}_{ij}$ is determined by 
\begin{eqnarray}
\label{ranksystem}
{\rm rank} \; G_1^{(2)} 
\!&\!=\!&\! {\rm rank} \; \Phi_{11} + {\rm rank}\; G_1^{(1)} 
= {\rm rank} \; \Phi_{12} + {\rm rank}\; G_2^{(1)}, \nonumber \\
{\rm rank} \; G_2^{(2)} 
\!&\!=\!&\! {\rm rank} \; \Phi_{21} + {\rm rank}\; G_1^{(1)}
= {\rm rank} \; \Phi_{22} + {\rm rank}\; G_2^{(1)}. 
\end{eqnarray}
For (\ref{DMV}), the difference in rank between consecutive 
symmetries is $1,$ so ${\rm rank} \; \Phi_{ij} = 1, \; i,j = 1,2. $
\vskip 5pt
\noindent
(i) As for the scalar case, we first construct the differential operator 
${\bf \Phi}_0.$ 
In view of the weights (\ref{weightsDMV}), 
\begin{equation}
\label{permDMV}
{\cal L}_{ij} = \{ u^2, v^2, u v, {\rm D}, \beta, \delta, \theta \}. 
\end{equation}
Hence, 
\begin{equation}
\label{diffopDMV}
{\bf \Phi}_0 \!=\!
\left(\!\!\!
\begin{array}{cc}
c_1 u^2 \!+\! c_2 v^2 \!+\! c_3 u v \!+\! c_4 {\rm D} \!+\! c_5 & 
c_6 u^2 \!+\! c_7 v^2 \!+\! c_8 u v \!+\! c_9 {\rm D} \!+\! c_{10} \\
c_{11} u^2 \!+\! c_{12} v^2 \!+\! c_{13} u v \!+\!c_{14} {\rm D}\!+\!c_{15}& 
c_{16} u^2 \!+\! c_{17} v^2 \!+\! c_{18} u v \!+\! c_{19} {\rm D} \!+\!c_{20}
\end{array}
\!\!\!\right),
\end{equation} 
where $c_5, c_{10}, c_{15}$, and $c_{20}$ will be linear in 
$\beta, \delta,$ and $\theta.$
\vskip 5pt
\noindent
(ii) Using the conserved densities $\rho^{(1)}=u, \rho^{(2)}=v,$ 
and $\rho^{(3)} = u^2 + v^2,$ we have
\begin{eqnarray}
\label{inner1DMV}
{\rm D}_t \rho^{(1)}\!&\!=\!&\!{\rm D}_t u = 
\frac{\partial u}{\partial u} F_1 + \frac{\partial u}{\partial v} F_2  
= (1, 0) \cdot (u_t, v_t) = - {\rm D}_x J^{(1)}, \\
{\rm D}_t \rho^{(2)}\!&\!=\!&\!{\rm D}_t v = 
\frac{\partial v}{\partial u} F_1 + \frac{\partial v}{\partial v} F_2  
= (0, 1) \cdot (u_t, v_t) = - {\rm D}_x J^{(2)}, \\
{\rm D}_t \rho^{(3)}\!&\!=\!&\!{\rm D}_t (u^2 + v^2) =
\frac{\partial (u^2 + v^2)}{\partial u} F_1 + 
\frac{\partial (u^2 + v^2) }{\partial v} F_2  \nonumber \\
\!&\!=\!&\!2 (u, v) \cdot (u_t, v_t) = - {\rm D}_x J^{(3)}, 
\end{eqnarray}
where the dot $(\cdot)$ refers to the standard inner product of vectors.
Therefore, introducing the symmetry ${(u_x, v_x)}^{\rm T}$ on the left 
of ${\rm D}^{-1}$ gives
\begin{equation}
\label{kforDMV} 
{\cal M} = \left \{ 
{(u_x, v_x)}^{\rm T} \odot {\rm D}^{-1} (u, v) 
\right \}, 
\end{equation}
where $\odot$ stands for the tensor product of matrices and ${\rm T}$
for transpose. So, 
\begin{eqnarray}
\label{phi1forDMV}
{\bf \Phi}_1\!&\!=\!&\!
c_{21}\; {(u_x, v_x)}^{\rm T} \odot {\rm D}^{-1} (u, v)
\nonumber \\
\!&\!=\!&\! \left (
   \begin{array}{cc}
    c_{21} u_x {\rm D}^{-1} u  & c_{21} u_x {\rm D}^{-1} v \\
    c_{21} v_x {\rm D}^{-1} u & c_{21} v_x {\rm D}^{-1} v 
   \end{array}
\right). 
\end{eqnarray}
Note that ${(u_x, v_x)}^{\rm T} \odot {\rm D}^{-1} (1,0)$ and 
${(u_x, v_x)}^{\rm T} \odot {\rm D}^{-1} (0,1)$ are of rank 
$\frac{1}{2}.$ They cannot be used in ${\bf \Phi}_1, $
where all pieces must have rank $1.$  

To determine the unknown constants in 
${\bf \Phi} = {\bf \Phi}_0 + {\bf \Phi}_1$, we use
\begin{equation}
\label{DMVrecopcond}
{\bf \Phi} {\bf G}^{(k)} = {\bf G}^{(k+1)} 
+ \sum_{l = 1}^{k} \alpha_{kl} {\bf G}^{(l)}, 
\quad k = 1, 2, \ldots,
\end{equation}
where $\alpha_{kl}$ are unknown coefficients (which can be zero).
In contrast to the examples in the previous sections, 
the $\alpha_{kl}$ play a role when dealing with systems with 
weighted parameters like (\ref{DMV}). 

It suffices to take $k =1$ and $2$ in (\ref{DMVrecopcond}) to fix all 
coefficients $c_i,$ and the extra unknowns $\alpha_{11}, \alpha_{21}$ and 
$\alpha_{22}.$ By solving a linear system, we obtain $\alpha_{11}=0, 
\alpha_{21} = \theta^2$ and $\alpha_{22} = \beta - \delta,$
and the values for the coefficients $c_1$ through $c_{21}.$ 
The recursion operator then follows readily: 
\begin{eqnarray}
\label{finopDMV}
\!\!\!\!\!\!\!\!\! {\bf \Phi} \!&=\!&\! \left (
   \begin{array}{cc}
    \beta-\delta + 2 u^2 +  2 u_x {\rm D}^{-1} u & 
    \theta + 2 u v  - {\rm D} + 2 u_x {\rm D}^{-1} v \\
    \theta +  2 u v + {\rm D} +  2 v_x {\rm D}^{-1} u & 
    2 v^2 + 2 v_x {\rm D}^{-1} v
   \end{array}
\right ).
\end{eqnarray}
The recursion operator for the case $\gamma=\theta=\delta=0$ was 
computed analytically in \cite{RWetal95}. 
\section[About the Integrability Package {\em InvariantsSymmetries.m}]{%
         About the Integrability Package InvariantsSymmetries.m}

We briefly describe our package {\it InvariantsSymmetries.m}, which 
automatically performs the computation of conservation laws 
(invariants) and symmetries based on the algorithms in Sections 3 and 4. 

Users must have access to {\it Mathematica} 3.0 \cite{Wolfram96}. 
The files for our package are available from {\it MathSource} 
\cite{UGandWH98c}. The package includes instructions for installation, 
on-line help, documentation, and built-in examples. 

After proper installation, it is advisable to first run our notebook 
(called {\it Examples}), which is accessible through the browser
as part of the Add-on Package {\it Integrability}. 
The interactive examples in the notebook will help familiarize the user
with the syntax of our functions (see also \cite{UGandWH98c}). 

To use the package as part of a new notebook, start {\it Mathematica} 
and type \verb|In[1]:= <<Integrability`| to read in the package. 
You will get the following message:
\begin{verbatim}
Loading init.m for Integrability from AddOns.
\end{verbatim}
The key functions for conservation laws and symmetries of PDEs are
{\verb|PDEInvariants|} and {\verb|PDESymmetries|}.
These functions take the following arguments: the equations in the 
system, the dependent and independent variables, and the range for the rank. 

For example (\ref{DMV}), the first two lines below define the system
(\ref{DMV}). 
The third line produces the densities (\ref{DMVdens3})--(\ref{DMVdens5}). 
The fourth line gives the symmetries (\ref{DMVsymm1})--(\ref{DMVsymm3}).
\begin{verbatim}
In[2]:= pde1 := D[u[x,t],t]+D[u[x,t]*(u[x,t]^2+v[x,t]^2)+
        beta*u[x,t]+gamma*v[x,t]-D[v[x,t],x],x] == 0;

In[3]:= pde2 := D[v[x,t],t]+D[v[x,t]*(u[x,t]^2+v[x,t]^2)+
        theta*u[x,t]+delta*v[x,t]+D[u[x,t],x],x] == 0;

In[4]:= PDEInvariants[{pde1,pde2}, {u,v}, {x,t}, {1,3}];

In[5]:= PDESymmetries[{pde1,pde2}, {u,v}, {x,t}, {3/2,5/2}];
\end{verbatim}
Help about the functions and their options is available on-line.
For instance, type \verb|In[6]:= ??PDEInvariants| to obtain 
the function description. Part of it reads: 
\vskip 0.1pt
\noindent
\begin{verbatim}
PDEInvariants[eqn, u, {x,t}, R, opts] finds the invariant with rank R 
of a partial differential equation for the function u.
PDEInvariants[{eqn1,eqn2,...}, {u1,u2,...}, {x,t}, {Rmin,Rmax}, opts] 
finds the invariants with rank Rmin through Rmax.
x is understood as the space variable and t as the time variable.
\end{verbatim}
Typing \verb|In[7]:= ??PDESymmetries| produces descriptions like: 
\vskip 0.1pt
\noindent
\begin{verbatim}
PDESymmetries[eqn, u, {x,t}, R, opts] 
finds the symmetry with rank R of a partial differential equation 
for the function u.
\end{verbatim}
Information about the options of PDESymmetries is obtained by typing 
\vskip 1pt
\noindent
{\verb|In[8]:= ??WeightedParameters|}. It returns:
\begin{verbatim}
WeightedParameters is an option that determines the parameters with 
weight. If WeightedParameters -> {p1,p2,...}, then p1, p2, ... are 
considered as constant parameters with weight. 
The default is WeightedParameters -> {}.
\end{verbatim}
The option {\verb|WeightedParameters|} is useful when working with systems 
that lack uniformity in rank. In such cases, our software tries to resolve 
the problem by itself and prints appropriate messages.
When unsuccessful, the program will suggest the use of the 
{\verb|WeightedParameters|} option. 
Therefore, the option {\verb|WeightedParameters|} should not be used
until it is explicitly recommended by the software. 

Rules for the weights of variables can be entered via the option
{\verb|WeightRules|}: 
\begin{verbatim}
WeightRules is an option that determines the rules for weights of 
the variables. If WeightRules -> {Weight[u] -> val,...}, then scaling 
properties are determined under these rules. There is a built in 
checking mechanism to see if the given rules cause inconsistency.
\end{verbatim}
For PDEs, the \verb|MaxExplicitDependency| option allows one to compute
conserved densities or symmetries that explicitly depend on the
independent variables:
\begin{verbatim}
MaxExplicitDependency is an option for finding the invariant and 
generalized symmetries of PDEs and DDEs. 
If MaxExplicitDependency -> Max_Integer, then the program allows for 
explicit dependency of independent variables of maximum degree Max. 
The default is MaxExplicitDependency -> 0.
\end{verbatim}
\section{Software Review}

In this section, we briefly review software for the computation 
of conservation laws, higher-order symmetries and recursion operators.
In Table~\ref{HGtable3}, we give a summary and contact information. 

Higher-order symmetries can be computed with prolongation methods, and 
numerous software packages are available that can aid in the tedious 
computations inherent to such methods.
A 50 page survey of software for Lie symmetry computations, 
including generalized symmetries, can be found in \cite{WH96}, and 
a short update in \cite{WH97}. We will not repeat these software
reviews here. A survey of packages for conservation laws was first given 
in \cite{UGandWH97a}. 
However, to keep this chapter self-contained, we present a summary of 
that survey.  

Based on dilation invariance, Ito's programs in REDUCE (see 
\cite{MI86,MI94,MIandFK85}) compute polynomial higher-order 
symmetries and conserved densities for systems of evolution equations 
that are uniform in rank (no weighted parameters can be introduced).
Ito's latest program, called {\it SYMCD}, cannot be used to compute 
symmetries and densities that depend explicitly on the 
independent variables $t$ and $x$, nor can it handle systems 
with parameters. More details are given in \cite{UGandWH97a}. 

In \cite{BFetal97,BFetal87}, Fuchssteiner et al.\ present
algorithms to compute higher-order {\it symmetries} of evolution equations. 
Their algorithm in \cite{BFetal97} is based on Lie-algebraic techniques 
and uses commutator algebra on the Lie algebra of vector fields. 
Their approach is different from the usual prolongation method in that
no determining equations are solved. 
Instead, all necessary generators of the finitely generated Virasoro 
algebra are computed from one given element by direct Lie-algebraic
methods. Their code is available in MuPAD. 
In \cite{BFetal87}, Fuchssteiner et al.\ give code
to verify that recursion operators are hereditary.  
In \cite{BFetal97}, it is shown how to compute {\it mastersymmetries} 
from which the recursion operators can be retrieved.

The REDUCE program {\it FS} for ``formal symmetries'' was written by 
Gerdt and Zharkov \cite{VGandAZ90} (see also \cite{VG93,VG96}).
{\it FS} computes higher-order symmetries and conservation laws of 
polynomial type. The algorithm requires that the evolution equations 
be of order two or higher in the spatial variable.
However, this approach does not require that the evolution
equations be uniform in rank. 
With {\it FS}, one cannot compute symmetries that depend explicitly on 
the independent variables $t$ and $x.$ 
Applied to equations with parameters, {\it FS} computes
the conditions on the parameters using the symmetry approach. 

The PC package {\it DELiA}, written in Turbo Pascal by Bocharov 
\cite{AB91} and co-workers, is a commercial computer algebra system 
for investigating differential equations using Lie's approach.
The program deals with higher-order symmetries, conservation laws, 
integrability and equivalence problems. 
It has a special routine for systems of evolution equations.
The program requires the presence of second- or higher-order spatial 
derivative terms in all equations.
For systems with parameters, {\it DELiA} does not automatically 
compute the densities and symmetries corresponding to the 
(necessary) conditions on the parameters. 
One has to use {\it DELiA}'s integrability test first, 
to determine the conditions. Once the parameters are fixed, 
one can compute the densities and symmetries. 

Sanders and Wang have Maple and FORM code for the computation of 
symmetries in the scalar case, allowing zero and negative weights
\cite{JSandJW98a,JSandJW98b,JSandJW98c} and nonpolynomial
equations and symmetries.
This code relies on the Maple package {\it diffalg} \cite{FBetal95}
to do the reductions of solutions of ODEs (PDEs). 
See \cite{JSandJW97a} for theoretical foundations of the 
computation of conservation laws, and \cite{JSandJW97b} for the use 
of their algorithms in the integrability classification of KdV-type 
higher order PDEs. 

Wolf et al. \cite{TWetal98} have three packages, called 
{\it CONLAW 1/2/3}, in REDUCE for the computation of conservation laws.
There is no limitation on the number of independent variables.  
The approach uses Wolf's program {\it CRACK} for solving overdetermined 
systems of PDEs (see \cite{WH96,WH97}).
Wolf's algorithm is particularly efficient for showing the non-existence 
of conservation laws of high order. 
In contrast to our program, it also allows one to compute nonpolynomial 
conservation laws.

Hickman \cite{MH98} at the University of Canterbury, 
Christchurch, New Zealand has implemented a slight variation 
of our algorithm for conserved densities in Maple. 
Instead of computing the differential monomials in the density by 
repeated differentiation, Hickman uses a tree structure combining 
the appropriately weighted building blocks. 

\section{Conclusions}

The {\it Mathematica} package {\it InvariantsSymmetries.m} presented 
in this chapter can be used for computer-aided integrability detection 
of systems of nonlinear PDEs as they occur in various branches of
science and engineering. 

More precisely, our package is a tool to search for the first 
half a dozen conservation laws and symmetries. 
If our programs succeed in finding a large set of independent 
conservation laws or symmetries, there is a good chance that the system 
has infinitely many of these quantities. 
For instance, if the number of conservation laws is 4 or less, 
most likely the system is not integrable---at least not in its current
coordinate representation.

Applied to a system with parameters, our package can determine the 
conditions on the parameters so that the system admits a sequence 
of conserved densities or generalized symmetries. 

An actual proof of integrability, by showing the existence of an infinity 
of conservations laws or symmetries, must be done analytically 
(see \cite{JSandJW98b} for results in this direction).
On the other hand, constructing the recursion operator, and showing that 
it indeed satisfies the defining equation, provides conclusive proof of 
integrability.


\section*{Acknowledgements}

We acknowledge helpful discussions with Professors Jan Sanders and
Frank Verheest, and thank them for careful reading of the manuscript.

This research project was supported in part by the National 
Science Foundation of the United States of America under Grant CCR-9625421.

\newpage

\begingroup \font\prq=cmtt10 \prq \obeyspaces \hsize=7.5in
\parindent=0pt 
\begin{tabular}{||l|l|l|l||}  \hline \hline 
\multicolumn{4}{||c||}
{\rule[-3mm]{0mm}{8mm}
{\bf Table 3$\;\;\;$List of Software and Contact Information} } \\ \hline 
& & & \\ 
Name \& System & Scope & Developer(s) \& Address & Email Address \\
& & & \\ \hline 
& & & \\ 
CONLAW 1/2/3 & Conservation & T.$\!$ Wolf {\it et al.$\!$} & 
T.Wolf@maths.qmw.ac.uk \\
(REDUCE) & Laws & School of Math.$\!$ Sci.$\!$ & \\
& & Queen Mary \& &  \\
& & Westfield College & \\
& & University of London & \\
& & London E1 4NS, U.K. & \\ 
& & & \\  \hline
& & &  \\ 
DELiA & Conservation & A.$\!$ Bocharov {\it et al.$\!$} & 
alexeib@saltire.com \\
(Pascal) & Laws and & Saltire Software & \\
& Generalized & P.O. Box 1565 &  \\
& Symmetries & Beaverton, OR 97075 & \\
& & U.S.A. & \\ 
& & & \\  \hline
& & &  \\ 
FS & Conservation & V.$\!$ Gerdt \& A.$\!$ Zharkov & 
gerdt@jinr.dubna.su \\
(REDUCE) & Laws and & Laboratory of Computing & \\
& Generalized & Techniques \& Automation  & \\
& Symmetries & Joint Institute for & \\
& & Nuclear Research & \\
& & 141980 Dubna, Russia & \\
& & & \\  \hline
& & &  \\ 
Invariants & Conservation & \"{U}.$\!$ G\"{o}kta\c{s} \& W.$\!$ Hereman & 
unalg@wolfram.com \\
Symmetries.m & Laws and & 
Dept.$\!$ of Math.$\!$ Comp.$\!$ Sci.$\!$ & whereman@mines.edu \\
(Mathematica) & Generalized & Colorado School of Mines & \\ 
& Symmetries & Golden, CO 80401, U.S.A.  & \\
& & &  \\ \hline \hline
\end{tabular}
\vfil
\newpage
\begin{tabular}{||l|l|l|l||} \hline \hline 
\multicolumn{4}{||c||}
{\rule[-3mm]{0mm}{8mm}
{\bf Table 3 cont.$\;\;\;$List of Software and Contact Information}}\\ \hline 
& & & \\ 
Name \& System & Scope & Developer(s) \& Address & Email Address \\
& & & \\ \hline 
& & & \\ 
SYMCD & Conservation & M.$\!$ Ito & ito@puramis.amath. \\
(REDUCE) & Laws and & Dept.$\!$ of Appl.$\!$ Maths.$\!$ & 
hiroshima-u.ac.jp \\ 
& Generalized & Hiroshima University & \\
& Symmetries & Higashi-Hiroshima & \\
& & 724 Japan & \\
& & & \\  \hline
& & &  \\ 
symmetry \& & Generalized & B.$\!$ Fuchssteiner {\it et al.$\!$} &
benno@uni-paderborn.de \\
mastersymmetry & Symmetries & Dept.$\!$ of Mathematics & \\
(MuPAD) & & Univ.$\!$ of Paderborn  & \\
& & D-33098 Paderborn & \\
& & Germany & \\
& & & \\  \hline
& & &  \\ 
Tests for & Conservation & J.$\!$ Sanders \& J.$\!$P.$\!$ Wang & 
jansa@cs.vu.nl \\
Integrability & Laws, Genera-& 
Dept.$\!$ of Math.$\!$ & \\ 
(Maple \& FORM) & lized Symmetries, & \& Comp.$\!$ Sci.$\!$ & \\
& and Recursion & Vrije Universiteit & \\ 
& Operators & 1081 HV Amsterdam & \\
& & The Netherlands & \\
& & & \\  \hline
& & &  \\ 
Tools for & Conservation & M.$\!$ Hickman & M.$\!$Hickman \\
Conservation Laws & Laws & Dept.$\!$ of Maths.$\!$ \& Stats.$\!$ & 
@math.canterbury.ac.nz\\ 
(Maple) & & University of Canterbury & \\ 
& & Private Bag 4800 & \\
& & Christchurch & \\
& & New Zealand & \\
& & &  \\ \hline \hline
\end{tabular}
\endgroup





\begin{thebibliography}{99}

\bibitem{MAandPC91}[1]
M. J. Ablowitz and P. A. Clarkson,
{\em Solitons, Nonlinear Evolution Equations and Inverse Scattering},
Cambridge University Press, Cambridge, 1991.

\bibitem{AB93}[2]
A. H. Bilge, 
On the equivalence of linearization and formal symmetries as integrability
tests for evolution equations, 
Journal of Physics A: Mathematics and General {\bf 26} (1993) 7511--7519.

\bibitem{AB91}[3]
A. V. Bocharov,
DELiA: a system for exact analysis of Differential Equations using
S. Lie Approach, DELiA 1.5.1 User Guide, 
Beaver Soft Programming Team, New York, 1991.

\bibitem{FBetal95}[4]
F. Boulier, D. Lazard, F. Ollivier and M. Petitot,
Representation for the radical of a finitely generated differential ideal, 
in: {\it Proceedings of ISSAC '95 International Symposium
on Symbolic and Algebraic Computation}, Ed.: A. H. M. Levelt, 
ACM Press, New York, 1995, pp. 158--166.

\bibitem{BDetal93a}[5]
B. Deconinck, P. Meuris and F. Verheest, 
Oblique nonlinear Alfv\'en waves in strongly magnetized beam plasmas. 
Part 1. Nonlinear vector evolution equation,
Journal of Plasma Physics {\bf 50} (1993) 445--455.

\bibitem{BDetal93b}[6]
B. Deconinck, P. Meuris and F. Verheest, 
Oblique nonlinear Alfv\'en waves in strongly magnetized beam plasmas. 
Part 2. Soliton solutions and integrability,
Journal of Plasma Physics {\bf 50} (1993) 457--476.

\bibitem{AF87}[7]
A. S. Fokas, 
Symmetries and integrability,
Studies in Applied Mathematics {\bf 77} (1987) 253--299.

\bibitem{BFetal97}[8]
B. Fuchssteiner, S. Ivanov and W. Wiwianka,
Algorithmic determination of infinite-dimensional symmetry groups 
for integrable systems in 1+1 dimensions,
Mathematical and Computer Modelling {\bf 25} (1997) 91--100.

\bibitem{BFandWO1982}[9]
B. Fuchssteiner and W. Oevel,
The bi-Hamiltonian structure of some nonlinear fifth- and seventh-order
differential equations and recursion formulas for their symmetries and
conserved covariants,
Journal of Mathematical Physics {\bf 23} (1982) 358--363.

\bibitem{BFetal87}[10]
B. Fuchssteiner, W. Oevel and W. Wiwianka,
Computer-algebra methods for investigation of hereditary operators of 
higher order soliton equations,
Computer Physics Communications {\bf 44} (1987) 47--55.

\bibitem{VG93}[11]
V. P. Gerdt,
Computer algebra, symmetry analysis and integrability of nonlinear
evolution equations,
International Journal of Modern Physics C {\bf 4} (1993) 279--286.

\bibitem{VG96}[12]
V. P. Gerdt,
Homogeneity of integrability conditions for multi-parametric families
of polynomial non-linear evolution equations,
Mathematics and Computers in Simulation {\bf 42} (1996) 399--408.

\bibitem{VGandAZ90}[13]
V. P. Gerdt and A. Y. Zharkov,
Computer generation of necessary integrability conditions for
polynomial-nonlinear evolution systems,
in: {\it Proceedings of ISSAC '90 International Symposium on 
Symbolic and Algebraic Computation}, 
Eds.: S. Watenabe and M. Nagata, 
ACM Press, New York, 1990, pp. 250--254.

\bibitem{UG98}[14]
\"{U}. G\"{o}kta\c{s},
{\em Algorithmic Computation of Symmetries, Invariants and Recursion Operators
for Systems of Nonlinear Evolution Equations and Differential-difference 
Equations}, 
Ph.D. Thesis, Department of Mathematical and Computer Sciences, 
Colorado School of Mines, Golden, Colorado, 1998. 

\bibitem{UGandWH97a}[15]
\"{U}. G\"{o}kta\c{s} and W. Hereman,
Symbolic computation of conserved densities for systems of nonlinear
evolution equations,
Journal of Symbolic Computation {\rm 24} (1997) 591--621.

\bibitem{UGandWH98a}[16]
\"{U}. G\"{o}kta\c{s} and W. Hereman,
Computation of conservation laws for nonlinear lattices, 
Physica D {\rm 123} (1998) 425--436.

\bibitem{UGandWH98b}[17]
\"{U}. G\"{o}kta\c{s} and W. Hereman,
Algorithmic computation of generalized symmetries for nonlinear 
evolution and lattice equations, to appear in 
Advances in Computational Mathematics (1999).

\bibitem{UGandWH98c}[18]
\"{U}.\ G\"{o}kta\c{s} and W.\ Hereman,
The {\it Mathematica} package {\it InvariantsSymmetries.m} is available 
at http://www.mathsource.com/cgi-bin/msitem?0208-932.
{\it MathSource} is the electronic library of 
Wolfram Research, Inc., Champaign, Illinois, 1998.

\bibitem{UGetal97}[19]
\"{U}. G\"{o}kta\c{s}, W. Hereman and G. Erdmann,
Computation of conserved densities for systems of nonlinear
differential-difference equations, 
Physics Letters A {\rm 236} (1997) 30--38.

\bibitem{BGandAR97}[20]
B. Grammaticos and A. Ramani, 
Integrability -- and how to detect it, 
in: Integrability of Nonlinear Systems, 
Eds.: Y. Kosmann-Schwarzbach, B. Grammaticos and K. M. Tamizhmani,
Springer Verlag, Berlin, 1997, pp. 30--94. 

\bibitem{WH96}[21]
W. Hereman,
Symbolic software for Lie symmetry analysis, 
in: CRC Handbook of Lie Group Analysis of Differential Equations, 
Volume 3: New Trends in Theoretical Developments and Computational Methods, 
Ed.: N. H. Ibragimov, CRC Press, Boca Raton, Florida, 1996, pp. 367--413. 

\bibitem{WH97}[22]
W. Hereman,
Review of symbolic software for Lie symmetry analysis,
Mathematical and Computer Modelling {\bf 25} (1997) 115--132.

\bibitem{WHetal98}[23]
W. Hereman, \"{U}. G\"{o}kta\c{s}, M. Colagrosso and A. Miller,
Algorithmic integrability tests for nonlinear differential and
lattice equations, 
Computer Physics Communications {\bf 115} (1998) 428--446. 

\bibitem{MH98}[24]
M. Hickman, Private communication, 1998.

\bibitem{MI86}[25]
M. Ito,
A REDUCE program for finding symmetries of nonlinear evolution 
equations with uniform rank,
Computer Physics Communications {\bf 42} (1986) 351--357. 

\bibitem{MI94}[26]
M. Ito, 
SYMCD - a REDUCE package for finding symmetries and conserved densities
of systems of nonlinear evolution equations,
Computer Physics Communications {\bf 79} (1994) 547--554.

\bibitem{MIandFK85}[27]
M. Ito and F. Kako,
A REDUCE program for finding conserved densities of partial differential
equations with uniform rank,
Computer Physics Communications {\bf 38} (1985) 415--419.

\bibitem{AMetal91}[28]
A. V. Mikhailov, A. B. Shabat and V. V. Sokolov,
The symmetry approach to classification of integrable equations, 
in: {\em What Is Integrability?}, V. E. Zakharov ed., 
Springer-Verlag, Berlin Heidelberg, 1991, pp. 115--184.

\bibitem{RMetal68}[29]
R. M. Miura, C. S. Gardner and M. D. Kruskal, 
Korteweg-de Vries equations and generalizations. II. Existence of
conservation laws and constants of motion, 
Journal Mathematical Physics {\bf 9} (1968) 1204--1209.

\bibitem{Olver93}[30]
P. J. Olver, 
{\em Applications of Lie Groups to Differential Equations,} 
2nd Edition, Springer Verlag, New York, 1993.

\bibitem{VRandGK94}[31]
V. Rosenhaus and G. H. Katzin,
On symmetries, conservation laws, and variational problems for partial
differential equations,
Journal of Mathematical Physics {\bf 35} (1994) 1998--2012.

\bibitem{JSandJW97a}[32]
J. A. Sanders and J. P. Wang,
Hodge decomposition and conservation laws,
Mathematics and Computers in Simulation {\bf 44} (1997) 483--493.

\bibitem{JSandJW97b}[33]
J. A. Sanders and J. P. Wang,
Classification of conservation laws for KdV-like equations,
Mathematics and Computers in Simulation {\bf 44} (1997) 471--481.

\bibitem{JSandJW97c}[34]
J. A. Sanders and J. P. Wang, 
On hereditary recursion operators,
Report WS-472, Department of Mathematics and Computer Sciences, 
Vrije Universiteit of Amsterdam, Amsterdam, The Netherlands, 1997.

\bibitem{JSandJW98a}[35]
J. A. Sanders and J. P. Wang,
On the classification of integrable systems, 
in: {\it Proceeding of the Fourth International Conference on 
Mathematical and Numerical Aspects of Wave Propagation}, 
Ed.: J. A. DeSanto, SIAM, Philadelphia (1998) pp. 393--397.

\bibitem{JSandJW98b}[36]
J. A. Sanders and J. P. Wang,
On the integrability of homogeneous scalar evolution equations,
Journal of Differential Equations {\bf 147} (1998) 410--434.

\bibitem{JSandJW98c}[37] 
J. A. Sanders and J. P. Wang,
Combining Maple and Form to decide on integrability questions,
Computer Physics Communications {\rm 115} (1998) 447-459.

\bibitem{KSandTK74}[38]
K. Sawada and T. Kotera,
A method for finding N-Soliton solutions of the K.d.V. equation and
K.d.V.-like equation, 
Progress in Theoretical Physics {\bf 51} (1974) 1355--1367.

\bibitem{JPW98}[39]
J. P. Wang, 
{\it Symmetries and Conservation Laws of Evolution Equations}, 
Ph.D. Thesis, Department of Mathematics and Computer Sciences, 
Vrije Universiteit of Amsterdam, Amsterdam, The Netherlands, 1998.

\bibitem{RWetal95}[40]
R. Willox, W. Hereman and F. Verheest,
Complete integrability of a modified vector 
derivative nonlinear Schr{\"o}dinger equation,
Physica Scripta {\bf 52} (1995) 21--26. 

\bibitem{TWetal98}[41]
T. Wolf, A. Brand and M. Mohammadzadeh,
Computer algebra algorithms and routines for the computation of
conservation laws and fixing of gauge in differential expressions,
Journal of Symbolic Computation {\bf 27} (1999) 221-238.

\bibitem{Wolfram96}[42]
S. Wolfram,
{\em The Mathematica book}, 3rd Edition,
Wolfram Media, Urbana-Champaign, Illinois 
\& Cambridge University Press, London, 1996.

\end{thebibliography}
\end{document}